\documentclass{article}
\usepackage{amscd,amsmath,amssymb,bbm}
\usepackage{amsfonts}
\newcommand{\be}{\begin{equation}}
\newcommand{\ee}{\end{equation}}
\newcommand{\bea}{\begin{eqnarray}}
\newcommand{\eea}{\end{eqnarray}}
\newcommand{\ba}{\begin{array}}
\newcommand{\ea}{\end{array}}

\newcommand{\nn}{\nonumber}
\newcommand{\cN}{{\cal N}}

\newcommand{\R}{{\mathbb{R}}}
\newcommand{\unity}{\mathbbm{1}}

\def\sfrac#1#2{{\textstyle\frac{#1}{#2}}}
\def\pa{\partial}

\def\={\ =\ }

\def\vph{$\vphantom{\big|}$}
\def\ax{\alpha{\cdot}x}

\def\bx{\beta{\cdot}x}
\def\hx{\hat{x}}
\def\hX{\hat{X}}
\def\+{${+}1$}
\def\-{${-}1$}
\def\0{${}\ 0$}

\def\a{\alpha}
\def\b{\beta}

\def\de{\delta}
\def\eps{\epsilon}

\begin{document}
\thispagestyle{empty}
\vspace{2cm}
\begin{flushright}
ITP--UH--11/09
\end{flushright}
\vspace{2cm}

\begin{center}
{\LARGE\bf A new class of solutions to the WDVV equation}
\vspace{1cm}

{\Large
Olaf Lechtenfeld$\,{}^{a}$ \ \ and \ \
Kirill Polovnikov$\,{}^{b}$
}
\vspace{1cm}

${}^a$
{\it  Institut f\"ur Theoretische Physik, Leibniz Universit\"at Hannover,
30167 Hannover, Germany}
\vspace{0.2cm}

${}^b$
{\it  Laboratory of Mathematical Physics, Tomsk Polytechnic University,
634050 Tomsk, Russia \ }
\end{center}
\vspace{3cm}

{\large
\begin{center}{\bf Abstract}\end{center}
\begin{center}\begin{minipage}{14cm}
\vph
The known prepotential solutions~$F$ to the Witten-Dijkgraaf-Verlinde-Verlinde
(WDVV) equation are parametrized by a set $\{\alpha\}$ of covectors.
\vph 
This set may be taken to be indecomposable, since
$F_{\{\alpha\}\smash{\dot\oplus}\{\beta\}}=F_{\{\alpha\}}+F_{\{\beta\}}$.
\vph
We couple mutually orthogonal covector sets 
\vph
by adding so-called radial terms to the standard form of~$F$.
\vph 
The resulting reducible covector set yields
a new type of irreducible solution to the WDVV equation.
\vph
\end{minipage}\end{center}
}

\newpage
\setcounter{page}{1}

\setcounter{equation}0
\section{Introduction}

The Witten-Dijkgraaf-Verlinde-Verlinde (WDVV) equation
\be\label{wdvv1}
F_i\ F_k^{-1} F_j \= F_j\ F_k^{-1} F_i 
\qquad\textrm{with}\quad i,j,k = 1,\ldots,n
\ee
is a nonlinear constraint on a set of $n$ $n{\times}n$ matrices $F_i$
whose entries are the third derivatives of a function $F(x^1,\ldots,x^n)$
called the prepotential,
\be
(F_i)_{pq} \= \pa_i \pa_p \pa_q F
\ee
(we abbreviate $\pa_i=\frac{\pa}{\pa x^i}$).
This equation was first introduced in the context of two dimensional
topological field theory \cite{wit,dvv} and $\cN{=}2$~SUSY
Yang-Mills theory \cite{marsh} and was intensively studied during
the following years. In 1999, it was shown~\cite{mart} that one can
construct solutions by taking the ansatz~\footnote{
The covector $\a$ evaluates on $x\equiv(x^1,\ldots,x^n)$ via
$\a(x)=\a_ix^i=:\ax$. Positive coefficients $f_\a$ may be absorbed in~$\a$.
}
\be\label{Fansatz}
F \= -\sfrac12\smash{\sum_{\a}} f_\a\ (\ax)^2\,\ln|\ax| \ ,
\ee
where $\{\a\}$ is the (positive) root system of some simple Lie algebra
of rank~$n$. This ansatz defines a constant metric given by the matrix
\be
G\= -x^i F_i \= \smash{\sum_\a} f_\a\ \a\otimes\a\ .
\ee
Shortly thereafter, it was proved~\cite{vesel1} that certain deformations
of such root systems still solve~(\ref{wdvv1}), and the list of possible 
collections of covectors~$\{\a\}$ was extended to the so-called 
$\vee$-systems~\cite{vesel2}.

In 2005, an interesting connection between the WDVV equation and
$\cN{=}4$~superconfermal mechanics was discovered~\cite{bgl}. 
In this context, a slightly different but equivalent formulation
of the WDVV equation appeared, namely
\be\label{wdvv2}
[\,F_i\,,\,F_j\,]\=0 \qquad\longleftrightarrow\qquad
(\pa_i\pa_k\pa_pF)(\pa_j\pa_l\pa_pF)\=(\pa_j\pa_k\pa_pF)(\pa_i\pa_l\pa_pF)\ ,
\ee
supplemented by the homogeneity condition
\be\label{homog}
-x^i F_i \= \unity \qquad\longleftrightarrow\qquad
-x^i \pa_i\pa_j\pa_k F\=\de_{jk}\ .
\ee
The latter picks a Euclidean metric in~$\R^n$ and determines a linear
change of coordinates which relates the two formulations within 
the ansatz~(\ref{Fansatz}) by mapping $G\mapsto\unity$~\cite{ioseph60}. 
We also write $x_j=\de_{jk}x^k=x^j$.

Rather recently, it was observed that the ansatz~(\ref{Fansatz}) can
be successfully extended by adding a `radial term'~\cite{glp},
\be\label{Fansatz2}
F\=-\sfrac12\smash{\sum_\a}f_\a\ (\ax)^2\,\ln|\ax|\ -\ \sfrac12f_R\,R^2\ln R^2
\ ,
\ee
where $\{\a\}$ is some $\vee$-system and $R$ denotes the radial coordinate
in~$\R^n$,
\be
R^2\ :=\ \smash{\sum_{i=1}^{n}}x_i^2 \qquad\textrm{so that}\qquad
G(x,x)\=\smash{\sum_{\a}}f_\a\ (\ax)^2\=(1{-}2f_R)\,R^2\ .
\ee
The two types of term in~(\ref{Fansatz2}) are extremal cases of a general
ansatz employing arbitrary quadratic forms.
For $n{>}2$ only two choices for $f_R$ are compatible with the WDVV equation,
namely $f_R=0$ and $f_R=1$, which are related by flipping the sign of the
metric~$G$ via $f_{\a}\mapsto-f_{\a}$.

In this Letter we further generalize the ansatz~(\ref{Fansatz2}) 
for the prepotential~$F$ and construct novel solutions to the WDVV equation.

\setcounter{equation}0
\section{Relative radial terms}
Let us consider a reducible covector system 
$L = L_1 \oplus L_2 \oplus \ldots \oplus L_M$,
which is a direct sum of $M$ irreducible covector collections with dimensions
$n_1,n_2,\ldots, n_M$, respectively. Each subsystem $L_I \ (I=1,\ldots,M)$
is chosen to solve the WDVV equation via the ansatz~(\ref{Fansatz}) for
$\a\in L_I$. Furthermore, we work in the `Euclidean coordinates' conforming 
to~(\ref{homog}). The prototypical example is the root system of a non-simple
but semi-simple Lie algebra. According to the covector set decomposition, 
we split the index set
\be
\{1,2,\ldots,n\} \= \{1,\ldots,n_1\}\cup\{n_1{+}1,\ldots,n_1{+}n_2\}\cup\cdots
\cup\{n{-}n_M{+}1,\ldots,n\} \= S_1\cup S_2\cup\cdots\cup S_M
\ee
and introduce the notation
\be
X^I_i = \begin{cases}
x_i & \textrm{if}\ \ i\in S_I \\[2pt] 0 & \textrm{otherwise}\end{cases}\ ,\quad
X^{IJ}_i = \begin{cases}
x_i & \textrm{if}\ \ i\in S_{IJ} \\[2pt] 0 & \textrm{otherwise}\end{cases}\quad 
\textrm{etc.}\qquad\textrm{for}\qquad S_{IJ}=S_I\cup S_J \quad\textrm{etc.}\ ,
\ee
and likewise for $\de^I_{ij}$, \ $\de^{IJ}_{ij}=\de^I_{ij}{+}\de^J_{ij}$ \ 
etc.. An important concept is that of {\it relative\/} radial coordinates
\be\label{relrad}
r_I^2\=\sum_{i\in S_I} x_i^2 \= x^i X^I_i\ ,\qquad
r_{IJ}^2\=r_I^2+r_J^2\=\!\sum_{i\in S_{IJ}} x_i^2 \= x^i X^{IJ}_i\ ,\qquad
\ldots\ ,\qquad R^2\=\sum_i x_i^2
\ee
for the subspaces 
$L_I$, \ $L_{IJ}=L_I{\oplus}L_J$, \ $L_{IJK}=L_I{\oplus}L_J{\oplus}L_K$ \ etc., 
all the way up to~$L$.

The key idea is to couple the mutually orthogonal components of this
reducible covector system by adding to the ansatz~(\ref{Fansatz}) not only
the overall radial term as in~(\ref{Fansatz2}) but also all possible
{\it relative\/} radial terms,
\be\label{f2}
F\= -\sfrac12\sum_{\a\in L}f_{\alpha}\,(\ax)^2\ln|\ax|
\ -\ \sfrac12\sum_{I=1}^M f_{r_I}\,r_I^2\,\ln r_I^2
\ -\ \sfrac12\sum_{I<J}^M f_{r_{IJ}}\,r_{IJ}^2\,\ln r_{IJ}^2\
\ -\ \ldots\ -\ \sfrac12\,f_R\,R^2\,\ln R^2 \ ,
\ee
defining a hierarchy 
$f_{r_I}\subset f_{r_{IJ}}\subset f_{r_{IJK}}\subset\ldots\subset f_R$
of radial couplings. We point out that this ansatz is no longer decomposable.

In order to verify~(\ref{f2}), we have to compute the WDVV coefficients
\bea
\pa_i\pa_j\pa_k F \= 
-\sum_{I=1}^M \sum_{\a\in L_I}f_{\a}\,\frac{\a_i\a_j\a_k}{\ax}
&-& 2 \sum_{I=1}^M f_{r_I}\Bigl(
\sfrac{X^I_i \de_{jk}^I + X^I_j \de_{ki}^I + X^I_k \de_{ij}^I}{r_I^2} -
\sfrac{2\, X^I_i X^I_j X^I_k}{r_I^4} \Bigr) \nn \\[4pt]
&-& 2\sum_{I<J}^M f_{r_{IJ}} \Bigl( \sfrac{
X^{IJ}_i\de_{jk}^{IJ}+X^{IJ}_j\de_{ki}^{IJ}+X^{IJ}_k\de_{ij}^{IJ}}{r_{IJ}^2} -
\sfrac{2\,X^{IJ}_i X^{IJ}_j X^{IJ}_k}{r_{IJ}^4} \Bigr) \\[4pt]
&-& \ldots\ \ -\ \ 2\,f_R\,\Bigl( 
\sfrac{x_i \de_{jk} + x_j \de_{ki} + x_k \de_{ij}}{R^2} -
\sfrac{2\,x_i x_j x_k}{R^4} \Bigr) \ . \nn
\eea
Contracting with $x^i$ should reproduce~(\ref{homog}).
Taking into account~(\ref{relrad}), one obtains a  system of equations,
\be\label{hom1}
\sum_{\a\in L_I} f_{\a}\,\a_i\a_j \ +\ 2\ \delta_{ij}^I
\Bigl( f_{r_I}\ + \sum_{J(\neq I)}^M f_{r_{IJ}}\ +
\sum_{J,K(\neq I)}^M f_{r_{IJK}} +\ \ldots\ +\ f_R \Bigr)
\= \delta_{ij}^I\ .
\ee
Since all covector collections $L_I$ are $\vee$-systems, 
we must have~\footnote{
This is not true for $n_I\le2$, because then $f_R$ may take any value
since the WDVV equation is empty.}
\be\label{epsi}
\smash{\sum_{\a\in L_I}} f_{\a}\,\a_i\a_j \= \eps_I\,\delta_{ij}^I
\qquad\textrm{with}\qquad \eps_I\in\{+1,-1\}\ , 
\ee
hence
\be\label{hom2}
f_{r_I}\ + \sum_{J(\neq I)}^M f_{r_{IJ}}\ + \sum_{J,K(\neq I)}^M f_{r_{IJK}}
\ +\ \ldots\ +\ f_R \= \begin{cases} 
0 & \textrm{for}\quad\eps_I=+1 \\[2pt] 1 & \textrm{for}\quad\eps_I=-1 
\end{cases}\ .
\ee
This takes care of the homogeneity condition~(\ref{homog}).
We come to the WDVV equation~(\ref{wdvv2}), which reads
\bea \label{FF}
&&\sfrac12 \sum_{\a,\b\in L}\!f_\a f_\b\,\frac{\a{\cdot}\b}{\ax\,\bx}\,
(\a\wedge\b)^{\otimes2}\ + \
\sum_{I=1}^M 4\,f_{r_I}\,\biggl\{1-f_{r_I}-2\,
\Bigl(\sum_{J(\neq I)}^M f_{r_{IJ}}{+}\ldots{+}f_R\Bigr)\biggr\}\,
\frac{T^I}{r_I^2} \\[6pt]
&&+\sum_{I<J}^M 4\,f_{r_{IJ}}\,\biggl\{1-f_{r_{IJ}}-2\,
\Bigl(\!\sum_{K(\neq I,J)}^M\! f_{r_{IJK}}{+}\ldots{+}f_R \Bigr)\biggr\}\,
\frac{T^{IJ}}{r_{IJ}^2} 
\ +\ \ldots\ +\ 4\,f_R\,\bigl\{1{-}f_R\bigr\}\,\frac{T}{R^2}\=0 \nn
\eea
with
\bea
&&(\a\wedge\b)^{\otimes2}_{ijkl}\=(\a_i\b_j-\a_j\b_i)(\a_k\b_l-\a_l\b_k)\ ,
\nn \\[6pt]
&&T_{ijkl}^I\=\de_{ik}^I\de_{jl}^I-\de_{il}^I\de_{jk}^I-\de_{ik}^I\hX_j^I\hX_l^I
+\de_{il}^I\hX_j^I\hX_k^I-\de_{jl}^I\hX_i^I\hX_k^I+\de_{jk}^I\hX_i^I\hX_l^I\ , 
\nn \\[6pt]
&&T_{ijkl}^{IJ}\=\de_{ik}^{IJ}\de_{jl}^{IJ}-\de_{il}^{IJ}\de_{jk}^{IJ}
-\de_{ik}^{IJ}\hX_j^{IJ}\hX_l^{IJ}+\de_{il}^{IJ}\hX_j^{IJ}\hX_k^{IJ}
-\de_{jl}^{IJ}\hX_i^{IJ}\hX_k^{IJ}+\de_{jk}^{IJ}\hX_i^{IJ}\hX_l^{IJ}\ , \\[6pt]
&& \qquad \qquad \qquad \cdots \nn \\[6pt]
&&T_{ijkl} \= \de_{ik}\de_{jl}-\de_{il}\de_{jk}-\de_{ik}\hx_j\hx_l
+\de_{il}\hx_j\hx_k-\de_{jl}\hx_i\hx_k+\de_{jk}\hx_i\hx_l\ ,\nn
\eea
where
\be
\hX_i^I\equiv\sfrac{X_i^I}{r_I}\ ,\qquad 
\hX_i^{IJ}\equiv\sfrac{X_i^{IJ}}{r_{IJ}}\ ,\qquad
\ldots\ ,\qquad \hx_i\equiv\sfrac{x_i}{R} \ .
\ee
Projecting onto the different (independent) poles in~(\ref{FF}), 
the WDVV equation requires that
\bea\label{fcoup}
&& \sum_{\a,\b\in L_I}\!f_\a f_\b\,\frac{\a{\cdot}\b}{\ax\,\bx}\,
(\a\wedge\b)^{\otimes2} \= 0 \ , \nn \\[6pt]
&& f_{r_I}\,\biggl\{1-f_{r_I}-2\,\Bigl(
\sum_{J(\neq I)}^M f_{r_{IJ}}{+}\ldots{+}f_R\Bigr)\biggr\} \= 0\ ,\\[6pt]
&& f_{r_{IJ}}\,\biggl\{1-f_{r_{IJ}}-2\,\Bigl(\!
\sum_{K(\neq I,J)}^M\! f_{r_{IJK}}{+}\ldots{+}f_R\Bigr)\biggr\}\=0\ ,\nn\\[6pt]
&& \qquad \qquad \qquad \cdots \nn \\[6pt]
&& f_R\,\bigl\{1{-}f_R\bigr\} \= 0 \ .\nn
\eea

Staring for a while at~(\ref{fcoup}) while taking into account~(\ref{hom2}),
one realizes that the only admissible radial couplings are
\be
f_{r_I},\ f_{r_{IJ}},\ f_{r_{IJK}},\ \ldots, f_R\ \in\{0,+1,-1\}
\qquad\textrm{but}\qquad f_R \neq-1\ .
\ee
The solutions are best described by a sequential procedure,
starting from the `radial-free' configuration
$f_{r_I}=f_{r_{IJ}}=\ldots=f_R=0\ \ \forall I,J,\ldots$,
corresponding to $\eps_I=+1\ \ \forall I$.
Now, let us turn on some radial couplings of the first hierarchy level,
$f_{r_I}=+1$ for some values of~$I$, which flips the signs of the
corresponding~$\eps_I$. On the next level, we may now switch on further
radial couplings~$f_{IJ}$, but only if they do not overlap.
For each nonzero $f_{IJ}$ we must flip the signs of the corresponding
$\eps_I$ and $\eps_J$ as well as those of $f_I$ and~$f_J$.
Continuing this scheme, we eventually arrive at the highest level,
where activating $f_R$ will flip all signs in the hierarchy.
In this way, a multitude of possible new WDVV solutions is generated.

\setcounter{equation}0
\section{Examples}
To illustrate the new possibilities for WDVV solutions, 
let us consider the semi-simple Lie algebra $A_1 \oplus A_2$.
Our generalized ansatz~(\ref{f2}) for this case ($M{=}2$) reads
\bea\label{example1}
F\!&=&\!-\sfrac12f_0\,(x_1{+}x_2{+}x_3)^2 \ln|x_1{+}x_2{+}x_3| \ -\
\sfrac12f\,\sum_{i<j}^3 (x_i{-}x_j)^2 \ln|x_i{-}x_j| \ -\
\sfrac12f_r\,r^2\ln r^2 \ - \ \sfrac12\,f_R\,R^2\ln R^2 \nn\\
&&\textrm{with}\qquad
r^2 \= \sfrac23(x_1^2{+}x_2^2{+}x_3^2{-}x_1x_2{-}x_1x_3{-}x_2x_3)
\qquad\textrm{and}\qquad R^2 \= x_1^2{+}x_2^2{+}x_3^2 \ .
\eea
The last term in~(\ref{example1}) couples the center of mass (the $A_1$ part)
to the relative motion (the $A_2$ part).
Since both subsystems are at most two-dimensional, the WDVV equation is
empty, being a consequence of the homogeneity condition~(\ref{homog}).
Hence, we only have to fulfill~(\ref{hom1}), which yields
\be
3f_0+2f_R=1 \quad\textrm{and}\quad 3f+3f_r+2f_R=1 
\qquad\longrightarrow\qquad f_R=\begin{cases} 
0 & \Rightarrow\quad f_0=f{+}f_r=+\sfrac13 \\[2pt] 
1 & \Rightarrow\quad f_0=f{+}f_r=-\sfrac13 \end{cases}\ ,
\ee
giving us a one-parameter family of solutions. \goodbreak

To see the power of the WDVV equation, 
we present a more generic $M{=}3$ example,
based on the Lie algebra $D_3 \oplus D_3 \oplus D_3$:
\bea
F\!&=&\!
-\sfrac12\,f_1\,\Bigl(\!\sum_{1\leq i<j\leq3}\!(x_i{+}x_j)^2 \ln|x_i{+}x_j|\
+\sum_{1\leq i<j\leq3}\!(x_i{-}x_j)^2 \ln|x_i{-}x_j| \Bigr)\ + \nn\\[6pt] &&
-\sfrac12\,f_2\,\Bigl(\!\sum_{4\leq i<j\leq6}\!(x_i{+}x_j)^2 \ln|x_i{+}x_j|\
+\sum_{4\leq i<j\leq6}\!(x_i{-}x_j)^2 \ln|x_i{-}x_j| \Bigr)\ + \nn\\[6pt] &&
-\sfrac12\,f_3\,\Bigl(\!\sum_{7\leq i<j\leq9}\!(x_i{+}x_j)^2 \ln|x_i{+}x_j|\
+\sum_{7\leq i<j\leq9}\!(x_i{-}x_j)^2 \ln|x_i{-}x_j| \Bigr)\ +    \\[6pt] &&
-\sfrac12\,f_{r_1}\,r_1^2\,\ln r_1^2\ \ \,-\ \ 
 \sfrac12\,f_{r_2}\,r_2^2\,\ln r_2^2\ \ \,-\ \ 
 \sfrac12\,f_{r_3}\,r_3^2\,\ln r_3^2\ \ + \nn\\[8pt] && 
-\sfrac12\,f_{r_{12}} r_{12}^2 \ln r_{12}^2\ -\
 \sfrac12\,f_{r_{13}} r_{13}^2 \ln r_{13}^2\ -\
 \sfrac12\,f_{r_{23}} r_{23}^2 \ln r_{23}^2\ -\
 \sfrac12\,f_R R^2 \ln R^2 \ , \nn
\eea
where $f_I=\sfrac14\eps_I$ and
\be
r_1^2 =\sum_{i=1}^3 x_i^2\ , \quad
r_2^2 =\sum_{i=4}^6 x_i^2\ , \quad
r_3^2 =\sum_{i=7}^9 x_i^2\ , \qquad
r_{ij}^2 = r_i^2 + r_j^2 \ , \qquad
R^2 = r_1^2 + r_2^2 + r_3^3 = \sum_{i=1}^9 x_i^2\ .
\ee
We list here (up to permutations) 
all possible radial coupling configurations:\\[12pt]
\begin{tabular}{p{3mm}p{3mm}p{3mm}|p{3mm}p{3mm}p{3mm}|p{3mm}p{3mm}p{3mm}|p{3mm}}
${}\,\eps_1$ & ${}\,\eps_2$ & ${}\,\eps_3$ & 
$f_{r_1}$ & $f_{r_2}$ & $f_{r_3}$ &
${}\!f_{r_{12}}$ & ${}\!f_{r_{13}}$ & ${}\!f_{r_{23}}$ & $f_R$ \\ \hline
\+ & \+ & \+ & \0 & \0 & \0 & \0 & \0 & \0 & \0 \\
\- & \+ & \+ & \+ & \0 & \0 & \0 & \0 & \0 & \0 \\
\- & \- & \+ & \+ & \+ & \0 & \0 & \0 & \0 & \0 \\
\- & \- & \- & \+ & \+ & \+ & \0 & \0 & \0 & \0 \\
\- & \- & \+ & \0 & \0 & \0 & \+ & \0 & \0 & \0 \\
\+ & \- & \+ & \- & \0 & \0 & \+ & \0 & \0 & \0 \\
\+ & \+ & \+ & \- & \- & \0 & \+ & \0 & \0 & \0 \\
\- & \- & \- & \0 & \0 & \+ & \+ & \0 & \0 & \0 \\
\+ & \- & \- & \- & \0 & \+ & \+ & \0 & \0 & \0 \\
\+ & \+ & \- & \- & \- & \+ & \+ & \0 & \0 & \0 
\end{tabular}
\qquad\qquad
\begin{tabular}{p{3mm}p{3mm}p{3mm}|p{3mm}p{3mm}p{3mm}|p{3mm}p{3mm}p{3mm}|p{3mm}}
${}\,\eps_1$ & ${}\,\eps_2$ & ${}\,\eps_3$ &
$f_{r_1}$ & $f_{r_2}$ & $f_{r_3}$ &
${}\!f_{r_{12}}$ & ${}\!f_{r_{13}}$ & ${}\!f_{r_{23}}$ & $f_R$ \\ \hline
\- & \- & \- & \0 & \0 & \0 & \0 & \0 & \0 & \+ \\
\+ & \- & \- & \- & \0 & \0 & \0 & \0 & \0 & \+ \\
\+ & \+ & \- & \- & \- & \0 & \0 & \0 & \0 & \+ \\
\+ & \+ & \+ & \- & \- & \- & \0 & \0 & \0 & \+ \\
\+ & \+ & \- & \0 & \0 & \0 & \- & \0 & \0 & \+ \\
\- & \+ & \- & \+ & \0 & \0 & \- & \0 & \0 & \+ \\
\- & \- & \- & \+ & \+ & \0 & \- & \0 & \0 & \+ \\
\+ & \+ & \+ & \0 & \0 & \- & \- & \0 & \0 & \+ \\
\- & \+ & \+ & \+ & \0 & \- & \- & \0 & \0 & \+ \\
\- & \- & \+ & \+ & \+ & \- & \- & \0 & \0 & \+
\end{tabular}
\vspace{0.5cm}

\noindent{\bf Acknowledgements}\\
\noindent
We thank S.~Krivonos for collaboration at various stages of this project.
K.P. is grateful to the Institut f\"ur Theoretische Physik at the Leibniz 
Universit\"at Hannover for hospitality.
The research was supported by RF Presidential grant NS-2553.2008.2, RFBR
grant 09-02-00078 and the Dynasty Foundation.

\end{document}